\documentclass[preprintnumbers, floatfix, letterpaper, twocolumn,aps,prd,epsfig,nofootinbib,natbib,longbibliography]{revtex4-1}
\usepackage{graphicx}
\usepackage{epstopdf}
\usepackage{latexsym}
\usepackage{amssymb}
\usepackage{amsmath}
\usepackage{color}
\usepackage{float}
\usepackage{mathrsfs}
\usepackage{multirow}
\usepackage{tabularx,booktabs}
\newcolumntype{Y}{>{\centering\arraybackslash}X}

\usepackage[center]{subfigure}
\usepackage{makecell}%To keep spacing of text in tables
\setcellgapes{4pt}
\begin{document}

%%%%%%%%%%%%%%%%%%%%%%%%%%%%%%%%%%%%%%%%%%%%%%%%%%%%%%%%%%%%%%%
  \renewcommand\arraystretch{2}
 \newcommand{\bq}{\begin{equation}}
 \newcommand{\eq}{\end{equation}}
 \newcommand{\bqn}{\begin{eqnarray}}
 \newcommand{\eqn}{\end{eqnarray}}
 \newcommand{\nb}{\nonumber}
 \newcommand{\lb}{\label}
 \newcommand{\cb}{\color{blue}}
    \newcommand{\cc}{\color{cyan}}
        \newcommand{\cm}{\color{magenta}}
\newcommand{\rc}{\rho^{\scriptscriptstyle{\mathrm{I}}}_c}
\newcommand{\rd}{\rho^{\scriptscriptstyle{\mathrm{II}}}_c}
\newcommand{\PRL}{Phys. Rev. Lett.}
\newcommand{\PL}{Phys. Lett.}
\newcommand{\PR}{Phys. Rev.}
\newcommand{\CQG}{Class. Quantum Grav.}
 %%%%%%%%%%%%%%%%%%%%%%%%%%%%%%%%%%%%%%%%%%%%%%%%%%%%%%%%%%%%%%%

\title{Singularities of plane gravitational  waves in Einstein's General Relativity}

\author{Tongzheng Wang$^{1, 2}$}
\email{tongzhengwang@126.com}
\author{Jared Fier$^{3}$}
\email{Jared$\_$Fier@baylor.edu}
\author{Bowen Li$^{3}$}
\email{Bowen$\_$Li@baylor.edu}
\author{Guoliang L\"u$^{2}$}
\email{guolianglv@xao.ac.cn}
\author{Zhaojun Wang$^{2}$}
\email{wzj@xju.edu.cn}
\author{Yumei Wu$^{1, 3}$}
\email{Yumei$\_$Wu@baylor.edu}
\author{Anzhong Wang$^{1, 3}$}
\email{Anzhong$\_$Wang@baylor.edu}
\affiliation{$^{1}$Institute for Advanced Physics $\&$ Mathematics,
Zhejiang University of Technology, Hangzhou, 310032, China\\
$^2$ School of Physical Science and Technology,
Xinjiang University, Urumqi, 830046, China\\
$^3$GCAP-CASPER, Department of Physics, Baylor University, Waco, TX, 76798-7316, USA}

\date{\today}

\begin{abstract}

Similar to the Schwarzschild coordinates for spherical black holes, the Baldwin, Jeffery and Rosen (BJR) coordinates for plane 
gravitational waves are often singular, and extensions beyond such singularities are necessary, before studying asymptotic 
properties of such spacetimes at the null infinity of the plane, on which the gravitational  waves propagate. The latter is closely 
related to the studies of memory effects and soft graviton theorems. In this paper, we point out that in the BJR coordinates all 
the spacetimes are singular physically at the focused point $u = u_s$, except for the two cases: (1)  $\alpha =1/2, \; \forall \; 
\chi_n$; and (2) $\alpha =1, \; \chi_i = 0\; (i = 1, 2, 3)$, where $\chi_n$ are the coefficients in the expansion $\chi \equiv 
\left[{\mbox{det}}\left(g_{ab}\right) \right]^{1/4} = \left(u- u_s\right)^{\alpha}\sum_{n = 0}^{\infty}\chi_n \left(u - u_s\right)^n$ with 
$\chi_0 \not= 0$, the constant $\alpha \in (0, 1]$ characterizes the strength of the singularities, and $g_{ab}$ denotes the reduced 
metric on the two-dimensional plane orthogonal to the propagation direction of the wave. Therefore,  the hypersurfaces $u= u_s$
already represent the boundaries of  such  spacetimes, and the null infinity does not belong to them. As a result, they cannot be 
used to study properties of plane gravitational waves at null infinities, including memory effects and soft graviton theorems.

\end{abstract}

\maketitle

\section{Introduction}
\renewcommand{\theequation}{1.\arabic{equation}} \setcounter{equation}{0}

The memory effects of gravitational waves (GWs) have been attracted lots of attention (see, for example,  \cite{BGY17,Zhang2017,Maluf2017,Andrzejewski2018} and references therein),
especially  after the recent observations of several  GWs emitted from remote binary systems of either black holes \cite{GW150914,GW151226,GW170104,GW170814}
or neutron stars \cite{GW170817}. (The detections of several more GWs were announced lately \cite{GWs18}). Such effects might be possibly detected by LISA \cite{Favata10} or even  
by current generation of detectors, such as  aLIGO and aVIRGO \cite{Lasky16}.
Recently, such investigations gained new momenta due to the close relations between asymptotically symmetric  theorems of soft gravitons and GW memory effects \cite{HPS16,Strominger16}.

The characteristic feature of these effects is the permanent displacement of a test particle  after a burst of a GW passes \cite{YZ74,BG85,Chris91,BD92,Thorne92,Harte13}.
In addition, the passage of the GW affects not only the position of the test particle, but also its velocity. In fact, the  change of the velocity of the particle is also permanent
\cite{Souriau73,BT87,BP88,GP89,zhang2018a}.

When far from the sources, the emitted GWs can be well approximated by plane GWs, a subject that has been extensively studied, including their nonlinear interactions
\cite{SKMHH,Griffiths16,Wang19}. The spacetimes for plane GWs can be cast in various forms, depending on the choice of the coordinates and gauge-fixing. One of them  was
originally due to Baldwin, Jeffery and Rosen (BJR) \cite{BJ,Rosen}. Despite its several attractive features, the system of the BJR coordinates is often singular within a finite width of  a wave,
and when studying the asymptotic behavior of the spacetime, extension beyond this singular surface is needed. In this paper, we point out
that there exist actually two kind of singularities in plane gravitational wave spacetimes, one represents coordinate singularities, which can be removed by 
proper coordinate transformations, and the other represents really spacetime singularities, and physical quantities, such as distortions of test particles, become 
infinitely large when such singularities are approaching. Therefore, in the latter these singularities already represent the boundaries of the spacetimes and extensions 
beyond them are not only  impossible but also  not needed.  Since gravitational memory effects and soft graviton theorems are closely related to the asymptotical behaviors of 
plane GW spacetimes, in the latter the spacetimes cannot be used to study such properties.

In general relativity (GR), there are powerful Hawking-Penrose theorems \cite{HE72}, from which one can see that spacetimes
with quite ``physically reasonable" conditions are singular. However,  the theorems did not tell the nature of the
singularities, and  Ellis and Schmidt  classified them into two different kinds, {\em spacetime curvature singularities} and {\em coordinate
singularities} \cite{ES77}. Spacetime curvature singularities are further divided into two sub-classes, {\em scalar curvature singularities} and
{\em non-scalar curvature singularities}. If any of the 14 scalars \cite{CW77}, constructed from the 4-dimensional Riemann tensor
$R^{\sigma}_{\mu\nu\lambda}$ and its derivatives,
is singular, then the spacetime is said singular, and the corresponding singularity is a scalar one. If none of these scalars
is singular, spacetimes can be still singular. In particular,  tidal forces and/or  distortions
(which are the double integrals  of the tidal forces), experienced by an observer, may become infinitely large
 \cite{Ori00}. This kind of singularities is usually referred to as non-scalar curvature singularities.

 In the spacetimes of plane GWs, all the 14 independent scalars vanish identically \cite{SKMHH,Griffiths16}, so in such spacetimes the singularities can be either non-scalar (but real spacetime) singularities or
 coordinate singularities. In this paper, we shall  clarify this important point, by studying tidal forces and distortions of freely falling observers. In particular, we find that the singularities
 can be in general characterized by
 \bq
\lb{1.2}
\chi(u) \equiv e^{-U(u)/2} = \left(u - u_s\right)^{\alpha} \hat\chi(u),
\eq
where the plane GWs are moving along the null direction of $u = $ Constant, $U(u)$ is defined in Eq.(\ref{2.1}), $\alpha > 0$, and $\hat\chi(u)$ is given by Eq.(\ref{hchi}) with  $\hat\chi(u_s) \not= 0$ \footnote{In wave mechanics, caustics often occur, and
the wave functions become discontinuous across the singularities, see,  for example,  \cite{Hor79} and references therein. However, the problem in GR is more subtle.
As shown in this paper, real spacetime singularities can occur, and such singularities actually represent the physical boundaries of the spacetimes.}. But, the Einstein vacuum field equations require $0 < \alpha \le 1$ (See the discussions
given in the next section). Then, we find that the tidal forces and distortions are finite across the singular surface $u = u_s$ only in two particular cases, in which we have  
 \bq
\lb{1.3}
(i) \; \alpha = \frac{1}{2},  \; \forall \; \chi_n, \quad {\mbox{or}} \quad (ii) \; \alpha = 1, \;  \chi_i = 0\; (i = 1, 2, 3), 
\eq
where $\chi_n$ are the coefficients appearing in the expansion (\ref{hchi}). 

Therefore, {\em all the plane GW spacetimes are singular physically    at the focused point  $u = u_s$, exceptions are  only  the ones with $\alpha = 1/2$ or $ 1$}.  
As a result, all the plane GW spacetimes  cannot be used to study memory effects and soft graviton theorems, except the ones
with $\alpha = 1/2, \; 1$, as only these spacetimes that  can be possibly extended to null infinity, whereby can memory effects and soft graviton theorems can be studied. 

It should be noted that, although the measurement of the physical states in the parameter space $\alpha \in (0, 1]$ is infinitesimal, there still exist an
 infinite number of solutions of the Einstein field equations, which satisfy the asymptotic properties of Eqs.(\ref{1.2}) and (\ref{1.3}) at the focusing point
 $u = u_s$. This can be seen clearly when one works in the Brinkmann coordinates  \cite{Brinkmann1925}, as to be shown below.

Specifically, the paper is organized as follows. In Sec. II we shall first give a brief review over the singularities appearing in the  BJR coordinates, and then study the  tidal forces and distortions felt by a typical class
of observers, whose movements are confined within the ($u, v$)-plane, and show explicitly  that  tidal forces and distortions of these observers are finite only in the two particular cases given by Eq.(\ref{1.3}).
 Since lots of studies of memory effects of GWs have been carried out in the Brinkmann coordinates, in Sec. III we consider the singular behavior of the hypersurface $u = u_s$ in the
 Brinkmann coordinates, and find the singular behavior of the function ${\cal{A}}(u)$, the only function that appears in the Brinkmann metric (\ref{MetricB}). The paper is ended in Sec. IV, in which we derive our main conclusions, and
 present some discussing remarks.

\begin{figure}
{
\includegraphics[width=8cm]{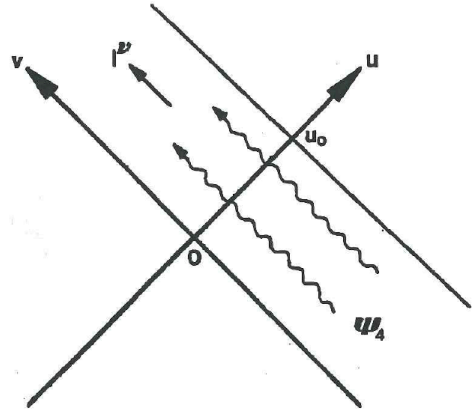}
 
}
\caption{A plane gravitational  wave moving along the null surfaces $u = $ const., with support only in the region $0 \le u \le u_0$, where $\Psi_4$ denotes the only non-vanishing component of the Weyl tensor, and is given by Eq.(\ref{Weyl})
in the  linearly polarized case.  }
\label{fig1}
\end{figure}

\section{Singularities in Spacetimes of Plane Gravitational Waves }
\renewcommand{\theequation}{2.\arabic{equation}} \setcounter{equation}{0}

The spacetimes for plane GWs   in the BJR coordinates  can be cast in the form \cite{SKMHH,Griffiths16,Wang19},
 \bqn
 \lb{2.1}
 ds^2 &=& -2e^{-M} du dv + e^{-U}\Big[e^{V}\cosh W dy^2 - 2\sinh W dydz \nb\\
 && + e^{-V}\cosh W dz^2\Big],
 \eqn
where  $M, U, V$ and $W$ are functions of $u$ only. The spacetime  in general represents a plane GW  moving  along the null surfaces $u = $ constant with two polarizations,
one is along the $y$-axis, often referred to as the ``+" polarization, and the other is along an axis which is at a $45^{o}$ degree with respect to the $y$-axis, often referred to as the 
``$\times$" polarization  [Cf. Fig.\ref{fig1}]. According to the Petrov classifications, the corresponding spacetimes belong to Petrov Type N  \cite{SKMHH,Griffiths16}. 

When the metric coefficients are functions of both $u$ and $v$, an interesting phenomenon raises, the gravitational  Faraday rotation, but now, the medium is provided by the nonlinear interaction of the
oppositely moving gravitational wave \cite{Wang19,Wang91th,Wang91}.

\subsection{Linearly Polarized Plane Gravitational Waves}

Note that by rescaling the null coordinate $u \rightarrow u' = \int{ e^{-M(u)} du}$, without loss of the generality,
one can always set
\bq
\lb{2.1a}
M = 0,
\eq
a gauge that will be adopted in this paper. In addition, for our current purpose, it is sufficient to consider only the linearly polarized case in which we have $W = 0$, so the metric
takes the simple form,
 \bqn
 \lb{MetricA}
 ds^2 &=& -2  du dv + e^{-U(u)}\left(e^{V(u)} dy^2 + e^{-V(u)} dz^2\right). ~~~~
 \eqn
 It can be shown that the corresponding Riemann tensor has only two independent components, given, respectively,  by
 \bqn
 \label{Riemann}
  R_{uyuy}&=&\frac{1}{4}e^{-(U-V)}\left[2\left(U''-V''\right) -(U'-V')^{2}\right], \nb\\
  R_{uzuz}&=&\frac{1}{4}e^{-(U+V)}\left[2\left(U''+V''\right) -(U'+V')^{2}\right], ~~~
\eqn
where $U' \equiv dU/d u$, etc. All the fourteen independent scalars \cite{CW77}, made of the Riemann tensor and its derivatives,  vanish identically \cite{SKMHH,Griffiths16}, so {\em there are no scalar singularities in the spacetimes of plane
GWs}.

Decomposing it into the Weyl and Ricci tensor \cite{SKMHH,Griffiths16}, we find that each of them has only one independent component. In particular, the independent component of
 the Ricci tensor is given by,
\bq
 \label{Ricci}
  R_{uu}= U'' - \frac{1}{2}\left({U'}^2 + {V'}^2\right),
\eq
while the   independent component of  the Weyl tensor is given by,
\bqn
 \label{Weyl}
 \Psi_4 \equiv - C_{\mu\nu\alpha\beta} n^{\mu} \bar{m}^{\nu} n^{\alpha}\bar{m}^{\beta}
 =  -\frac{1}{2}A^2\left(V'' - U'V' \right), ~~~~~
\eqn
which represents the plane GWs propagating along the hypersurfaces $u = $ const., as illustrated  in Fig.\ref{fig1}, where
 \bqn
 \lb{2.2}
 && l^{\mu} \equiv  A^{-1} \delta^{\mu}_{v}, \quad n^{\mu} \equiv A\delta^{\mu}_{u}, \quad m^{\mu} =  \zeta^2 \delta^{\mu}_2 +  \zeta^3 \delta^{\mu}_3,\nb\\
 &&
 \bar{m}^{\mu} = \overline{\zeta^2} \delta^{\mu}_2  + \overline{\zeta^3}  \delta^{\mu}_3,
 \eqn
form a null tetrad, with $A$ being an arbitrary function of $u$ only, and
 \bqn
 \lb{2.3}
 \zeta^2 \equiv \frac{e^{(U-V)/2}}{\sqrt{2}}, \quad
  \zeta^3 \equiv i \frac{e^{(U+V)/2}}{\sqrt{2}}.
 \eqn
An over bar denotes the complex conjugate. As noticed in various occasions,  the BJR coordinates are not harmonic,  typically not global, and contain coordinate singularities,
 see, for example, \cite{Rosen1937,Bondi1959,Zhang2017} and references therein.

To overcome these problems,  the Brinkmann coordinates $(\hat u, \hat v, \hat y, \hat z)$ are often used, defined by,
\bqn
\lb{2.4}
\hat v &\equiv& v+\frac{1}{4} y^2e^{V-U}\left(V' - U'\right) -\frac{1}{4} z^2e^{-V-U}\left(V' + U'\right),\nb\\
\hat u &\equiv&  u, \quad
\hat y \equiv e^{(V-U)/2} y, \quad
\hat z \equiv  e^{-(V+U)/2} z,
\eqn
in terms of which, the metric (\ref{MetricA}) takes the form  \cite{Brinkmann1925},
 \bqn
 \lb{MetricB}
 ds^2 &=& -2  d\hat u d\hat v + d\hat{y}^2  + d\hat{z}^2 + \frac{1}{2}{\cal{A}}(\hat{u})\left(\hat{y}^2 - \hat{z}^2\right) d\hat{u}^2,  ~~~~~~~~
 \eqn
where
 \bqn
 \lb{2.5}
 {\cal{A}}(\hat{u}) \equiv \frac{1}{2}\left[2\left(V'' - U''\right) +\left(V' - U'\right)^2\right].
 \eqn
 
 As we mentioned previously, in this paper we would like to point out that these singularities are not always coordinate ones. In fact, {\em all singularities are really spacetime singularities  at the focused point $u = u_s$, except only
 the ones that asymptotically behave as that given by Eqs.(\ref{1.2}) and (\ref{1.3}) at the neighborhood of the focused point}. To show our claim, we find that it is easier 
 to work in the BJR coordinates. Since the nature of singularities does not depend on the choice of coordinates, they must be singular in any   coordinate system, including the  Brinkmann system of coordinates.

\subsection{Spacetime Singularities}

In the vacuum case, the Einstein field equations $R_{\mu\nu} = 0$ have only one independent component, given by $R_{uu} = 0$, and from Eq.(\ref{Ricci}) we find that it can be written
as,
\bq
 \label{2.6}
\chi'' + \omega^2 \chi = 0,
\eq
where
\bq
\lb{2.7}
\chi \equiv e^{-U/2}, \quad \omega \equiv \frac{1}{2} V'.
\eq
Then, from Eq.(\ref{2.6}) we can see that, for any given initial value, $\chi_0 > 0$, there always exists a moment, say, $u = u_s$ at which $\chi$ vanishes \cite{Zhang2017},
\bq
\lb{2.8}
\chi(u_s) = 0, \quad {\mbox{or}} \quad U(u_s) = + \infty,
\eq
that is, a singularity of the metric (\ref{MetricA}) appears at $u = u_s$, which is surely not a scalar singularity, since, as mentioned above, all the fourteen independent scalars made
of the Riemann tensor in such spacetimes vanish identically. Dose this mean that the singularity must be a coordinate one? The answer is not always affirmative, 
as   non-scalar spacetime singularities can be present \cite{ES77}. In particular,  distortions of a freely falling observer,
which are the twice integrations of the tidal force with respect to the proper time of the observer, can diverge  \cite{Ori00}.

To calculate distortions of a freely falling observer, let us first consider the trajectory of the observer. In the present paper, we just consider the ones laid in the
($u, v$)-plane, that is, $(u, v, y, z) = (u(\lambda), v(\lambda), y_0, z_0)$, where $\lambda$ denotes the proper time of the observer, and $y_0$ and $z_0$ are constants. Then, the timelike geodesics are simply
given by
\bqn
\label{frame1}
u =\gamma_{0} \lambda, \quad v =\frac{\lambda}{2\gamma_{0}}, \quad y = y_0, \quad z = z_0,
\eqn
where  $\gamma_{0}$ is an integration constant. Define $e^{\mu}_{(0)}\equiv dx^{\mu}/d\lambda$, we can construct a tetrad, $e^{\mu}_{(a)}\; (a = 0, 1, 2, 3)$,  by
\bqn
\label{frame2}
  e^{\mu}_{(0)}&=& \gamma_{0}\delta^{\mu}_{u}+\frac{1}{2\gamma_{0}}\delta^{\mu}_{v}, \quad
  e^{\mu}_{(1)}=\gamma_{0}\delta^{\mu}_{u}-\frac{1}{2\gamma_{0}}\delta^{\mu}_{v},\nb\\
  e^{\mu}_{(2)}&=&e^{\frac{U-V}{2}}\delta^{\mu}_{y}, \quad
  e^{\mu}_{(3)}=e^{\frac{U+V}{2}}\delta^{\mu}_{z},
\eqn
which satisfies the relations,
\begin{equation}
\label{frame3}
  e^{\mu}_{(\alpha)}e^{\nu}_{(\beta)}g_{\mu\nu}=\eta_{\alpha\beta}, \quad
  e^{\mu}_{(\alpha);\nu}e^{\nu}_{(0)}=0,
\end{equation}
that is, they are unit orthogonal vectors and parallelly transported alone the timelike geodesics, so that they form a freely falling frame \cite{Ori00}. Then, the projection of
the Riemann tensor onto this frame, $R_{(a)(b)(c)(d)}\equiv R_{\mu\nu\lambda\rho}e^{\mu}_{(a)}e^{\nu}_{(b)}e^{\lambda}_{(c)}e^{\rho}_{(d)}$,
 yields two independent components,
 \bqn
 \label{RiemannB}
  R_{(0)(2)(0)(2)}&=&\gamma_0^2 e^{U-V}R_{uyuy}, \nb\\
  R_{(0)(3)(0)(3)} &=& \gamma_0^2 e^{U+V}R_{uzuz},
\eqn
where $R_{\mu\nu\lambda\rho}$'s are given by Eq.(\ref{Riemann}).

To study the nature of the singularities at $u = u_s$, we assume that in the neighborhood of $u= u_s$, the function $\chi$ takes the form,
 \bqn
 \label{chi}
\chi(u)  = \left(u - u_s\right)^{\alpha} \hat\chi(u), 
\eqn
where $\alpha > 0$ and $\hat\chi(u_s) \not=0$ but finite. Thus, expanding it as 
\bq
\lb{hchi}
\hat\chi(u) = \sum_{n= 0}^{\infty}{\chi_n (u-u_s)^n},
\eq
we must  assume that $\chi_0 \not= 0$, since $\hat\chi(u_s) \not= 0$. Then, from Eqs.(\ref{2.6}) and (\ref{2.7}) we find that
\bqn
\lb{2.20}
V' &=&  \left(- \frac{4\chi''}{\chi}\right)^{1/2} =  \frac{2 }{u-u_s}\Bigg[\alpha(1-\alpha)\nb\\
&&   - 2\alpha(u-u_s)\frac{\hat\chi'}{\hat\chi} - (u-u_s)^2 \frac{\hat\chi''}{\hat\chi}\Bigg]^{1/2},\nb\\
U &=& -2\ln\chi = -2\alpha\ln(u-u_s) - 2\ln\hat\chi(u).
\eqn
Note that, in writing the above expression for $V'$ we had chosen the plus sign, without loss of generality.
 To study the singular behavior of the solutions at the  focused point further, it is found convenient to consider the cases with and without $\alpha = 1$, separately.

\subsubsection{$\alpha = 1$}

In this case, inserting Eq.(\ref{hchi}) into Eq.(\ref{2.20}), we obtain  
\bqn
\lb{2.21}
V' &=& \frac{2\sqrt{-2\chi_1/\chi_0}}{\left(u-u_s\right)^{1/2}}\sum_{n=0}^{\infty}{v_n\left(u-u_s\right)^n},\nb\\
V'' &=& \frac{2\sqrt{-2\chi_1/\chi_0}}{\left(u-u_s\right)^{3/2}}\sum_{n=0}^{\infty}{\left(n - \frac{1}{2}\right)v_n\left(u-u_s\right)^n},\nb\\
U' &=& -\frac{2 }{u-u_s}\Bigg[ 1 + \frac{\chi_1}{\chi_0}\left(u-u_s\right) \nb\\
&& - \frac{\chi_1^2-2\chi_0\chi_2}{\chi_0^2}\left(u-u_s\right)^2 + ...\Bigg],
\eqn
where
\bqn
\lb{2.22}
v_0 &=& 1, \quad v_1 = - \frac{\chi_1^2 - 3\chi_0\chi_2}{2\chi_0\chi_1}, \nb\\
v_2 &=&   \frac{3 \chi_{1}^{4} - 10 \chi_{0} \chi_{1}^{2} \chi_{2} - 9 \chi_{0}^{2} \chi_{2}^{2} + 24 \chi_{0}^{2} \chi_{1} \chi_{3}}{8\chi_0^2\chi_1^2}, ~~~~~
\eqn
and $\chi_n$ are coefficients appearing in Eq.(\ref{hchi}). Hence, from Eqs.(\ref{Riemann}) and (\ref{RiemannB}) we find 
 \bqn
     \lb{2.23}
     R_{(0)(2)(0)(2)}  &=&-R_{(0)(3)(0)(3)}=\frac{1}{2}\gamma_{0}^{2}\left(U'V'-V''\right) \nb\\
     &=& -\frac{3 \sqrt{-\chi_{1}/(2\chi_{0})} \gamma_{0}^{2}}{(u - u_{s})^{3/2}}\nb\\
     && +\frac{3 (\chi_{1}^{2} + 5 \chi_{0} \chi_{2}) \gamma_{0}^{2}}{2\chi_{0}^{2} \sqrt{-2\chi_{1}/\chi_{0}} (u - u_{s})^{1/2} }\nb\\
     &&+{\cal{O}}\left((u-u_{s})^{\frac{1}{2}}\right), 
     \eqn
and   
\bqn
     \label{2.9b}
     && \int{d\lambda \int{d\lambda \; R_{(0)(2)(0)(2)}}}(\lambda) = 6\sqrt{-\frac{2\chi_{1}}{\chi_{0}}}\gamma_{0}^{\frac{1}{2}}(\lambda-\lambda_{s})^{\frac{1}{2}}\nb\\
      &&  + {\cal{O}}\left((\lambda- \lambda_s)^{\frac{3}{2}}\right),
     \eqn
which is finite as $\lambda \rightarrow \lambda_s$, where $\lambda_s \equiv u_s/\gamma_0$. 

\subsubsection{$\alpha \not= 1$}

In this case, we find that 
\bqn
\lb{2.24}
V' &=& \frac{2}{u-u_s}\sum_{n=0}^{\infty}{v_n\left(u-u_s\right)^n},\nb\\
V'' &=& \frac{2}{\left(u-u_s\right)^{2}}\sum_{n=0}^{\infty}{\left(n - 1\right)v_n\left(u-u_s\right)^n},\nb\\
U' &=& -\frac{2 }{u-u_s}\Bigg[\alpha  + \frac{\chi_1}{\chi_0}\left(u-u_s\right) \nb\\
&& - \frac{\chi_1^2-2\chi_0\chi_2}{\chi_0^2}\left(u-u_s\right)^2 + ...\Bigg],
\eqn
but now with
\bqn
\lb{2.25}
v_0 &=& \sqrt{(1-\alpha)\alpha}, \quad v_{1} = - \frac{\chi_{1}\alpha}{\chi_{0}\sqrt{(1-\alpha)\alpha}},   \nb\\
v_2 &=&  - \frac{\alpha[\chi_{1}^{2}\alpha(-1+2\alpha)+2\chi_{0}\chi_{2}(1+\alpha-2\alpha^2)]}{2\chi_{0}^{2}[(1-\alpha)\alpha]^{3/2}}. ~~~~~
\eqn
Clearly, to have the metric coefficient $V$ be real, we must assume that $0 < \alpha < 1$. Then, we find that 
   \bqn
     \lb{2.26}
     R_{(0)(2)(0)(2)} &=&-R_{(0)(3)(0)(3)}=\frac{1}{2}\gamma_{0}^{2}\left(U'V'-V''\right) \nb\\
     &=&\frac{\gamma_{0}^{2}(1-2\alpha)\sqrt{(1-\alpha)\alpha}}{(u-u_{s})^{2}}\nb\\
     && +\frac{2 \chi_{1} \alpha (2 \alpha -1) \gamma_{0}^{2}}{ \chi_{0} \sqrt{(1 - \alpha) \alpha}\left(u -u_{s}\right)} \nb\\
      &&- \frac{\alpha\gamma_0^2}{2 \chi_{0}^{2} [(1 - \alpha) \alpha]^{3/2}} \bigg[\chi_{1}^{2} \alpha \left(-7 + 12 \alpha - 8 \alpha^{2}\right)\nb\\
      && + 2 \chi_{0} \chi_{2} \left(-1 + \alpha - 8 \alpha^{2} + 8 \alpha^{3}\right)\bigg] \nb\\
       &&+{\cal{O}}\left(u-u_s\right).
     \eqn
Note that only the first term leads to divergency in the distortions. In fact,  we have 
 \bqn
 \label{2.9a}
&& \int{d\lambda \int{d\lambda \; R_{(0)(2)(0)(2)}}} = (2\alpha -1)\sqrt{\alpha(1-\alpha)}\nb\\
&& \times  \ln(\lambda- \lambda_s)   + {\cal{O}}\Big((\lambda- \lambda_s)ln(\lambda- \lambda_s)\Big),
  \eqn
for   $0<\alpha<1$. Clearly, they are  always singular, unless $\alpha = 1/2$. 

Combining the above with  the case $\alpha = 1$, we conclude that, {\em unless 
\bqn
 \label{NSs}
(i) \; \alpha = \frac{1}{2}, \quad {\mbox{or}} \quad (ii) \; \alpha = 1,
\eqn
the singularities located at the focused point $u = u_s$ are always really spacetime singularities}.

\section{Singularities in Brinkmann Coordinates}
\renewcommand{\theequation}{3.\arabic{equation}} \setcounter{equation}{0}

As mentioned previously, gravitational memory effects are frequently studied in the  Brinkmann coordinates. 
So, it would be very interesting to see how the metric behaves in the neighborhood of $u = u_s$ in the  Brinkmann coordinates.

From Eqs.(\ref{2.7}) and (\ref{chi}), we find that
\bqn
\lb{3.1}
U &=& - 2\alpha \ln\left(u - u_s\right) - 2 \ln \hat\chi(u),\nb\\
{V'}^2 &=& \frac{4\alpha(1-\alpha)}{(u-u_s)^2}
- \frac{4}{\hat\chi}\left(\hat\chi'' +  \frac{2\alpha\hat\chi'}{u - u_s}\right),
\eqn
where we can expand $\hat\chi(u)$ in the neighborhood of $u = u_s$ as that given by Eq.(\ref{hchi}).

In the vacuum, Eq.(\ref{2.6}) holds, from which we find that
\bq
\lb{3.3}
2U'' - {U'}^2 = {V'}^2, \; (R_{\mu\nu} = 0).
\eq
Then, Eq.(\ref{2.5}) reduces to,
 \bqn
 \lb{2.5b}
 {\cal{A}}(u) =  V''  -V'U', \; (R_{\mu\nu} = 0).
 \eqn
Note that in writing the above expression we used the coordinate transformations (\ref{2.4}), from which we simply find $u = \hat{u}$.
 Inserting Eqs.(\ref{3.1}) and (\ref{hchi})  into Eq.(\ref{2.5b}), we can find the behavior of ${\cal{A}}({u})$ in the neighborhood of $u = u_s$.
 
It is interesting to note that  tidal forces between two nearby null geodesics of $\hat{v}, \hat{y}, \hat{z} =$ Constant were studied
in the Brinkmann coordinates (\ref{MetricB}) in \cite{MZ03}, and showed that  ${R}_{\hat{y} \hat{u}\hat{y} \hat{u}}   
=  {\cal{A}}(\hat{u})/2$ 
describes diffeomorphism-invariant curvature information. Therefore, the divergence of $ {\cal{A}}(\hat{u})$ at the focusing point $u = u_s$ implies 
the existence of a spacetime singularity. In the following we shall show that this is consistent with our conclusions obtained in the last section for
the case $\alpha = 1/2$, and helps us understand the case $\alpha = 1$ in more details.

We find that it is convenient to consider the cases, (i) $0 < \alpha < 1, \; \alpha \not= 1/2$; (ii)  $ \alpha = 1/2$;  and  (iii) $\alpha  = 1$, separately.

\subsection{$0 < \alpha < 1, \; \alpha \not= \frac{1}{2}$}

 In this case, inserting Eqs.(\ref{3.1}) and (\ref{hchi}) into Eq.(\ref{2.5}), we find that
 \begin{equation}
 \label{3.4}
  {\cal{A}}(u) = \sum_{n = -2}^{\infty}{\mathscr{A}_{n} \left(u-u_{s}\right)^{n}},
\end{equation}
where the first three coefficients that show  the singular behavior of ${\cal{A}}(u)$ are given by,
\bqn
\label{A1A}
\mathscr{A}_{-2}&=& -2 (1- 2\alpha)\sqrt{(1-\alpha)\alpha},\nb\\
\mathscr{A}_{-1} &=& \frac{4\chi_{1}\alpha(1-2\alpha)}{\chi_{0}\sqrt{(1-\alpha)\alpha}},  \nb\\
\mathscr{A}_{0} &=& - \frac{\alpha}{[(1-\alpha)\alpha]^{3/2}\chi_{0}^{2}}\Big[\chi_{1}^{2}\alpha(7-12\alpha+8\alpha^{2}) \nb\\
&& +2\chi_{0}\chi_{2}(1-\alpha+8\alpha^{2}-8\alpha^{3})\Big].
\eqn

Since $0 < \alpha < 1$ and $\alpha \not= {1}/{2}$, we have $\mathscr{A}_{-2} \not= 0$, so  the leading divergent term now is $(u-u_s)^{-2}$, and  ${\cal{A}}(u)$ behaves as
\bqn
\label{3.6}
 {\cal{A}}(u) &=&  \frac{\mathscr{A}_{-2}(\alpha)}{\left(u-u_{s}\right)^{2}}  + \frac{{\mathscr{A}}_{-1}(\alpha)}{\left(u-u_{s}\right)} + \mathscr{A}_{0}\left(\chi_0, \chi_1, \chi_2\right)\nb\\
 &&  +  {\cal{O}}\left(u-u_{s}\right),
\eqn
in the neighborhood of $u = u_s$, where ${\mathscr{A}}_{-1}(\alpha)$ is a function of $\alpha$ only, which is also non-zero for $0 < \alpha < 1$ and $\alpha \not= {1}/{2}$, as it can be seen from
Eq.(\ref{A1A}). As mentioned in the last section, the spacetime now
is singular, and no extension beyond this surface is possible, so $u = u_s$ represents a real boundary of the spacetime.
This is consistent with the analysis in the Brinkmann coordinates  given in \cite{MZ03}. 

From Eq.(\ref{3.1}) we find that
\bqn
\lb{3.6a}
U(u) &=& - 2\alpha ln\left(u - u_s\right) + \hat{U}(u),\nb\\
V(u) &=& 2 \sqrt{\alpha(1-\alpha)}\; ln\left(u - u_s\right) + \hat{V}(u),
\eqn
where  $\hat{U}$ and $\hat{V}$ are regular and finite functions of $u$ across the hypersurface $u = u_s$. Note that in writing down the above expressions, we took the positive sign of $V'$, without loss of  generality,
as we did previously.  In addition,   $\hat{U}$ and $\hat{V}$ are not independent, as
they must satisfy the field equation (\ref{3.3}).

\subsection{ $\alpha=\frac{1}{2}$}

In this case, the singularity at $u = u_s$ is a coordinate singularity, which can be removed by the coordinate transformations of Eq.(\ref{2.4}), and the resulted metric is the Brinkmann metric (\ref{MetricB})
with
\begin{equation}
\label{B1/2}
  {\cal{A}}(u) =\sum_{n = 0}^{\infty}{\mathscr{B}_{n} \left(u-u_{s}\right)^{n}},
\end{equation}
where the first term ${\mathscr{B}_{0}}$ is given by, %
\bqn
\label{B1B}
\mathscr{B}_{0}&=&-\frac{6   (\chi_{1}^2 + 2 \chi_{0} \chi_{2})}{\chi_{0}^2}.
\eqn
Clearly, in this case ${\cal{A}}(u)$ is well-behaved in the neighborhood of $u = u_s$, and  the  Brinkmann metric (\ref{MetricB}) can be considered as its extension beyond the hypersurface $u = u_s$. If such obtained ${\cal{A}}(u)$ is analytical, then the extension is unique. Again, this is consistent with the analysis in the Brinkmann coordinates  presented in \cite{MZ03}.

On the other hand, from Eq.(\ref{3.1}) we find that
\bqn
\lb{3.6b}
U(u) &=& - ln\left(u - u_s\right) + \hat{U}(u),\nb\\
V(u) &=&  ln\left(u - u_s\right) + \hat{V}(u),\; \left(\alpha = {1}/{2}\right),
\eqn
where $\hat{U}$ and $\hat{V}$ are regular and finite functions of $u$ across the hypersurface $u = u_s$, and are related each other  through Eq.(\ref{3.3}).

\subsection{$\alpha=1$}

In this case, from Eq.(\ref{3.1}) we find that
\bqn
\lb{3.6c}
U &=& - 2 \ln\left(u - u_s\right) + \hat{U}(u),\nb\\
{V'}^2 &=&    - \left(\frac{\hat\chi'}{\hat\chi}\right) \frac{8}{u - u_s} - \frac{4\hat\chi''}{\hat\chi},
\eqn
where $\hat\chi$ takes the form of  Eq.(\ref{hchi}) with $\chi_0 \not= 0$. Thus, depending on values of $\chi_1,\; \chi_2$ and $\chi_3$, the function $ {\cal{A}}(u) $
can have different singular behaviors. Therefore, in the following, let us consider them separately.

\subsubsection{$\chi_{1}\neq0$}

In this case,  we find that
\bqn
\lb{3.6d}
U &=& - 2 \ln\left(u - u_s\right) + \hat{U}(u),\nb\\
V &=&   4\sqrt{2}   {\cal{D}}_1 \left(u - u_s\right)^{1/2} + {\cal{O}}\left(\left(u - u_s\right)^{3/2}\right),\nb\\
{\cal{A}} &=& \frac{1}{\left(u-u_{s}\right)^{3/2}} \sum_{n = 0}^{\infty}{\mathscr{C}_{n}(u-u_{s})^{n}},
\eqn
where   ${\cal{D}}_1 \equiv  \sqrt{-\chi_1/\chi_0}$,  $\hat{U}$ is regular and finite functions of $u$ across the hypersurface $u = u_s$, and the leading terms of $\mathscr{C}_{n}$ that show clearly the singular behavior of 
${\cal{A}}$ are given by,
\bqn
\label{C1C}
\mathscr{C}_{0} &=&3 \sqrt{2}  \; {\cal{D}}_1,\nb\\
\mathscr{C}_{1}&=& -\frac{3}{\sqrt{2} \;  \chi_{0}^{2} {\cal{D}}_1}   \left(\chi_{1}^{2} + 5 \chi_{0} \chi_{2}\right), \nb\\
\mathscr{C}_{2} &=& -\frac{3}{4\sqrt{2}\chi_{0}^{4}{\cal{D}}_1^{3}}   \left(9 \chi_{1}^{4}-14\chi_{0} \chi_{1}^{2}\chi_{2}+21\chi_{0}^{2}\chi_{2}^{2}\right.\nb\\
&& \left. -56\chi_{0}^{2}\chi_{1}\chi_{3}\right).
\eqn

It is interesting to note that in the current case the Brinkmann metric is still singular at $u = u_s$, although the distortions felt by the freely falling observers defined
by Eq.(\ref{frame1}) are all finite. Hence, now there are two possibilities: (i) Distortions felt by other freely falling observers diverge at $u = u_s$, so the singularity is a real spacetime singularity, and the spacetime cannot
be extended beyond this surface. (ii) Distortions felt by any of freely falling observers are finite, and the singularity is a coordinate one. 
The results given in \cite{MZ03} show that the current case belongs to the first possibility, as the tidal forces for two nearby null geodesics of $\hat{v}, \hat{y}, \hat{z} = $ Constant  become unbounded at $u = u_s$.

\subsubsection{$\chi_{1} = 0, \; \chi_{2} \not=0$}

When  $\chi_{1} = 0$ and $\chi_{2} \not=0$, we find that
\bqn
\lb{3.6e}
U &=& - 2 \ln\left(u - u_s\right) + \hat{U}(u),\nb\\
V &=&   2 \sqrt{6}   {\cal{D}}_2 \left(u - u_s\right) + {\cal{O}}\left(\left(u - u_s\right)^{2}\right),\nb\\
{\cal{A}} &=& \frac{1}{u-u_s}\sum_{n = 0}^{\infty}{\mathscr{D}_{n}(u-u_{s})^{n}},
\eqn
where  ${\cal{D}}_2 \equiv  \sqrt{-\chi_2/\chi_0}$,  $\hat{U}$ is regular and finite functions of $u$ across the hypersurface $u = u_s$, and the leading terms of $\mathscr{D}_{n}$ are given by,
\bqn
\label{D1D}
\mathscr{D}_{0}&=&4 \sqrt{6}   {\cal{D}}_2, \quad
\mathscr{D}_{1}= \frac{6 \sqrt{6}  {\cal{D}}_2 \chi_{3}}{\chi_{2}}, \nb\\
\mathscr{D}_{2} &=& -\frac{4   \sqrt{2/3} {\cal{D}}_2^{3}}{\chi_{2}^{3}} \big(3 \chi_{2}^{3} - 3 \chi_{0} \chi_{3}^{2} \nb\\
&& + 10 \chi_{0} \chi_{2} \chi_{4}\big).
 \eqn
Thus, now the Brinkmann metric is also singular near the hypersurface $u = u_s$, and the corresponding spacetimes are physically singular at $u = u_s$.

\subsubsection{ $\chi_{1} = \chi_{2}=0$, $\chi_3 \not= 0$}

In this case, we find that
\bqn
\lb{3.6f}
U &=& - 2 \ln\left(u - u_s\right) + \hat{U}(u),\nb\\
V &=&   \frac{8\sqrt{3}}{3}    {\cal{D}}_3 \left(u - u_s\right)^{3/2} + {\cal{O}}\left(\left(u - u_s\right)^{5/2}\right),\nb\\
 {\cal{A}} &=&  \frac{1}{(u-u_{s})^{\frac{1}{2}}} \sum_{n = 0}^{\infty}{\mathscr{E}_{n}(u-u_{s})^{n}},
\eqn
where  ${\cal{D}}_3 \equiv  \sqrt{-\chi_3/\chi_0}$ and 
\bq
\label{E1E}
\mathscr{E}_{0}=10 \sqrt{3}   {\cal{D}}_3, \quad
\mathscr{E}_{1}= \frac{35    {\cal{D}}_3 \chi_{4} }{ \sqrt{3} \chi_{3}}.
\eq
Again, in this case the Brinkmann metric is also singular, and a spacetime curvature singularity is developed on the focusing hypersurface  $u = u_s$.

\subsubsection{ $\chi_{1}=0 = \chi_{2}= \chi_{3}=0$, $\chi_4 \not= 0$}

 In this case, we find that
\bqn
\lb{3.6g}
U &=& - 2 \ln\left(u - u_s\right) + \hat{U}(u),\nb\\
V &=&  2\sqrt{5}   {\cal{D}}_4   \left(u - u_s\right)^{2} + {\cal{O}}\left(\left(u - u_s\right)^{3}\right),\nb\\
  {\cal{A}} &=& \sum_{n= 0}^{\infty}{\mathscr{G}_{n} \left(u-u_{s}\right)^{n}},
\eqn
where ${\cal{D}}_{4} \equiv \sqrt{-\chi_{4}/\chi_{0}}$ and $\mathscr{G}_{0}=12 \sqrt{5} {\cal{D}}_{4}$.
In this case, it is clear that the Brinkmann metric becomes non-singular, and Eq.(\ref{2.4}) represents an extension of the singular BJR metric
(\ref{MetricA}) beyond the hypersurface $u = u_s$. So, in this case it is sure that the singularity encountering in the BJR metric is a coordinate one, and the Brinkmann metric (\ref{MetricB}) is one
of its extensions. Note that the extension will be unique, if such obtained ${\cal{A}}(u)$ is analytical across $u = u_s$.

\subsection{Examples of ${\cal{A}}(u)$}

In the studies of gravitational wave memory effects, several interesting cases have been considered. For example, in \cite{Zhang2017,Gibbons1971}, the function ${\cal{A}}(u)$
was chosen as
\bq
\lb{3.7}
{\cal{A}}(u)=\frac{1}{2}\frac{d^{3}e^{-u^{2}}}{du^{3}} = 2u\left(3-2u^2\right) e^{-u^{2}}.
\eq
Once ${\cal{A}}(u)$ is given, we can solve Eqs.(\ref{2.5}) and (\ref{3.3}),
 \bqn
 \lb{3.7a}
&& 2\left(V'' - U''\right) +\left(V' - U'\right)^2 =  2{\cal{A}}(u), \\
\lb{3.7b}
&& 2U'' - {U'}^2 = {V'}^2,
\eqn
to find $U$ and $V$. However, due to the nonlinearity of these equations, usually it is difficult to find analytical solutions. In  \cite{Zhang2017} it was found numerically that
the singularity in the BJR coordinates happen at $u_s \simeq 0.593342$. From Eq.(\ref{3.7}) we can see that ${\cal{A}}(u)$ is finite and well-behaved in the neighborhood of this point. So, it must belong to either the case with $\alpha = 1/2$, or the case with $\alpha = 1$ and $\chi_{i} = 0 \; (i = 1, 2, 3)$. Due to the high nonlinearity between the BJR and Brinkmann coordinates, we find it is difficult to get a definite answer.
Some modified versions of the above example were considered in \cite{Maluf2017,zhang2018,zhang2018a}. 

Another example is the case with \cite{Andrzejewski2018}, 
\bq
\lb{3.7}
{\cal{A}}(u)=\frac{2}{\pi} \frac{\varepsilon^{3}}{(u^{2}+\varepsilon^{2})^{2}},
\eq
 where $\varepsilon$ is a constant. When   $\varepsilon$ is very small, the above expression gives rise to an impulse gravitational waves, recently
studied in \cite{zhang2018}. Clearly, in all of these models, ${\cal{A}}(u)$ is always finite and well-behaved across the singularity located at $u = u_s$  in the BJR coordinates. So, they all belong to
the non-singular cases, presented in the current paper.

\section{Conclusions and Discussing Remarks}
\renewcommand{\theequation}{4.\arabic{equation}} \setcounter{equation}{0}

 The memory effects of gravitational  waves are tightly related to the asymptotical properties of the spacetime at the future null infinity (see Ref. \cite{Favata10,zhang2018a,Andrzejewski2018} and references therein) \footnote{Here the null infinities are referred to the ones $u \rightarrow \infty$, that is, the null infinities on the ($u, v$)-plane, papendicular to the polarization plane of the plane gravitational waves.},
 and so are the soft gravitons and black holes \cite{HPS16,Strominger16}. However, it is well-known that in the BJR coordinates (\ref{MetricA}), the metric coefficients often become singular, and extensions beyond
 the singularities are needed before studying these important issues.

 In this paper, we have first pointed out that such extensions are not always possible, as some of these singularities are physically real singularities. In particular,  distortions experienced by freely falling observers in the
 ($u, v$)-plane can be divergent, and any objects trying across the singular surface will be destroyed  by these distortions [cf. Eq.(\ref{2.9a})]. As a result, in these cases the singularities actually represent  the boundaries of
 the spacetimes. In particular, if the metric coefficient $e^{-U}$ vanishes at the singularity $u = u_s$ as,
 \bq
 \lb{4.1}
 \chi \equiv e^{-U/2} = \left(u - u_s\right)^{\alpha}\hat\chi(u),
 \eq
 where $\alpha \in (0, 1]$, which is required for the metric coefficients to be real,  and $\hat\chi\left(u_s\right) \not= 0$, we found that distortions experienced by such freely falling observers {\em always diverge, unless $\alpha = 1/2$ or $\alpha = 1$}. Therefore, only in the cases
 where $\alpha = 1/2$ or $1$, the spacetimes at $u = u_s$ are possibly non-singular, and extensions of the spacetimes beyond this surface is needed, whereby we are able to study the memory effects of gravitational  waves
 and  soft gravitons and black holes.

 Coordinate transformations  from the BJR coordinates to the Brinkmann ones are carried out by Eq.(\ref{2.4}). It is interesting to note that in the Brinkmann coordinates there is only one unknown function
 ${\cal{A}}$, while in the BJR coordinates there are two, $U$ and $V$. However, the vacuum Einstein field equation (\ref{3.3}) relates $U$ to $V$, so finally there is only one independent component, too. In fact, for any given
 $V$, from Eq.(\ref{3.3}) one can find $U$, and then the function ${\cal{A}}$ is uniquely determined by Eq.(\ref{2.5b}). It is also interesting to note that the inverse is not unique, that is, for any given ${\cal{A}}(u)$,
 Eqs.(\ref{3.7a}) and (\ref{3.7b}) will have a family of solutions of the form, $U(u, u_1, u_2)$ and $V(u, v_1, v_2)$, where $u_i$'s and $v_i$'s are the integration constants.

 With the above in mind, we find that  ${\cal{A}}$ is finite and well-behaved across $u = u_s$ for $\alpha = 1/2$ for any given $\chi_n$ [cf. Eq.(\ref{B1/2})], 
 where $\chi_n$ are the expansion coefficients of $\hat\chi(u)$, given in Eq.(\ref{hchi}). However, in the case $\alpha = 1$, we found that ${\cal{A}}$ is finite and well-behaved across
 $u = u_s$ only when $\chi_1 = \chi_2 = \chi_3 = 0$. If any of these three coefficients is not zero, ${\cal{A}}(u)$ will be singular across $u = u_s$,
 although the distortions of the freely falling observers considered in this paper are finite. There are two possibilities for these cases: (i) The corresponding spacetimes are indeed singular, and distortions become unbounded
 across $u = u_s$ for other kinds of observers. (ii) The corresponding singularities are coordinate ones, but the proper coordinate transformations are not given by Eq.(\ref{2.4}), and instead they are given by somethings else. The results on the studies of the tidal forces between two nearby null geodesics  in the Brinkmann coordinates presented in \cite{MZ03} show that the possibility (i) is the right answer, as the tidal forces will diverge whenever  ${\cal{A}}(u)$ does. 
 Therefore, it is concluded that {\em only in the two cases given by Eq.(\ref{1.2}), the spacetimes are not singular at the focusing surface $u = u_s$, and extensions
 beyond it are needed, in order to obtain maximal spacetimes.}

Finally we note that our results are expected to be valid when both of the two polarizations exist, that is, $W \not= 0$ in Eq.(\ref{2.1}), although in the current paper we only considered the case  $W = 0$.

 \section*{Acknowledgements}

We would like to thank J. Oost for his earlier collaboration and  valuable comments and suggestions. We would like also to express our gratitude to
P. A. Horv\'athy for  valuable suggestions and comments.
 This work was supported in part by the National Natural Science Foundation of China under Nos. 11473024, 11363005, 11763007,
 11563008, 11365022, 11375153 and 11675145, and the XinJiang Science Fund for Distinguished Young Scholars under No. QN2016YX0049.

\end{document}